\documentclass[12pt]{article}
\newcommand{\ignore}[1]{}
\usepackage[english]{babel}
\usepackage{graphicx,rotating,epsfig}

\newcommand{\TeV}{\ensuremath{\mathrm{Te\kern -0.1em V}}}
\newcommand{\GeV}{\ensuremath{\mathrm{Ge\kern -0.1em V}}}
\newcommand{\MeV}{\ensuremath{\mathrm{Me\kern -0.1em V}}}
\def\GeVc2{\ensuremath{\mathrm{ Ge\kern -0.1em V }\kern -0.2em /c^2 }}

\newcommand{\MW}{\ensuremath{M_{\mathrm{ W }}}}
\newcommand{\GW}{\ensuremath{\Gamma_{\mathrm{ W }}}}

\newcommand{\RunI}{\hbox{Run-I}}
\newcommand{\RunII}{\hbox{Run-II}}

\begin{document}
\begin{center}
{\large FERMI NATIONAL ACCELERATOR LABORATORY}
\end{center}

\begin{flushright}
     TEVEWWG/WZ 2010/01 \\
      FERMILAB-TM-2460-E \\  
        CDF Note 10103 \\
       D0 Note 6041 \\[1mm]
   March 12, 2010\\
\end{flushright}

\begin{center}
{\LARGE \bf Combination of CDF and D0 Results \\
                  on the  Width of the $W$ boson\\}

\vfill

{\Large
The Tevatron Electroweak Working Group\footnote{
The Tevatron Electroweak Working group can be contacted at tev-ewwg@fnal.gov.\\
 \hspace*{0.20in} More information is available at {\tt http://tevewwg.fnal.gov}.}\\
for the CDF and D0 Collaborations}

\vfill
\vskip 1 cm

{\bf Abstract}

\end{center}

{
We summarize and combine  direct measurements of the width of the $W$ boson
in data collected by the Tevatron experiments CDF and D0
at Fermilab. Results from CDF and D0 {\RunI}
(1992-1995)  have been 
combined with  the CDF 200 pb$^{-1}$ results
 from the first period of {\RunII} (2001-2004) and the recent 1 fb$^{-1}$
 result in the electron channel from D0 (2002-2006). The
results are corrected  for any inconsistencies in parton
distribution functions and assumptions about electroweak parameters used in the different analyses.  The resulting
Tevatron average for the width of the $W$ boson
is $\GW = $ 2,046 $\pm 49~\MeV$.}

\vfill

\section{Introduction}

The CDF and D0 experiments    at the Tevatron $\overline{p}p$ collider,
 located at the Fermi National Accelerator
Laboratory, have directly measured the total decay width of the W boson, $\GW$, in both the
$e\nu$ and $\mu\nu$ decay modes.

Previous measurements of $\GW$ were published by CDF using
{\RunI}~\cite{GW-CDF-Ia,GW-CDF-Ib} and 
{\RunII}~\cite{GW-CDF-II} data and by D0~\cite{GW-D0-I} in {\RunI}. These earlier results
have been combined by this group and appear in reference~\cite{TEV08}. The present note  includes
a more recent
2009 D0  measurement using 1 fb$^{-1}$ of data from{ \RunII} in the
$e\nu$ decay mode~\cite{GW-D0-II} and supersedes the previous Tevatron combinations ~\cite{TEV08, MWGW-RunI-PRD,GW0508}.
The measurements are
combined using  the analytic BLUE method~\cite{Lyons:1988,
Valassi:2003}.  
This procedure takes  account of both    statistical and
systematic uncertainties as well as the correlations among them.

As in the July 2008 combined   analysis of the mass and width in reference~\cite{TEV08} and the 2009 analysis of the mass measurements~\cite{TEV09},  there are three significant changes relative to previous combinations of $W$ boson results by the TEVEWWG:

\begin{itemize}

\item The individual   measurement channels of $\GW$ for CDF  Run-Ia
and Run-Ib  are now combined for each run period using the BLUE
method to achieve a consistent statistical treatment of  all 
results.

\item  The older central values of $\GW$, based upon  very old parton distribution function (PDF) sets, are corrected to
use the  PDFs  from the  CTEQ6M~\cite{CTEQ} PDF set with uncertainty estimates from the CTEQ6M, CTEQ6.1M~\cite{CTEQ61M} and  MRST2003~\cite{MRS}   sets.
The new D0 Run-II measurement uses CTEQ6.1M while the CDF Run II measurement used CTEQ6M. The difference in the width extracted using these PDF sets is found to be  $\GW(CTEQ6.1M) - \GW(CTEQ6M) = 3 \pm 9$ MeV,
significantly less than the 20 MeV PDF systematic uncertainty, and therefore no correction is applied for this difference.

\item In addition, in this combined measurement we have revisited the uncertainty on $\GW$ resulting from changes in the assumed value of $\MW$  in different measurements. The revised  uncertainties are described when they are used in section \ref{combo}.

\end{itemize}

\section{New data from D0 on $\GW$}

The width, \GW\, of the $W$ boson   determined by D0 in Run-II~\cite{GW-D0-II}, using $W\rightarrow e \nu$
decays observed in 1 fb$^{-1}$ of data at $\sqrt{s}=1960$ GeV, 
is extracted from    fits to the transverse mass distribution in the range $100 \le  m_T \le  200$ ~GeV.
  The central value of  $\GW$ is  2,028 $\pm$ 72~ MeV.
  The measurement procedure differs somewhat from the 2009 D0 Run-II   measurement~\cite{MW-D0-RUN2} of \MW\ 
as the effects of hadrons recoiling against the $W$ boson  in the  measurement of $\GW$  are modeled using a library of recoil 
kinematics derived from detected $Z$ bosons~\cite{NIM} rather than through
simulation, as  in the measurement of the $W$ boson mass. The $\MW$ measurement used
the transverse mass range $65 \le m_T \le 90$ GeV, in addition to fits to the $p_T^e \ge  25$ GeV and $p_T^\nu \ge 25$ GeV distributions, and, as a result, the $W$ event samples used in the \GW\ and \MW\ measurements are effectively statistically  independent.  The two measurements share electron response, resolution and  models of detector efficiencies,  and the same production models and electroweak radiative corrections.
  
 Table \ref{table:err} summarizes the uncertainties on  the new  measurement  of $\GW$ by D0 and the correlation of sources of systematic uncertainties with previous results.

\begin{table}[!hbp]
\begin{center}
\begin{tabular}{|l |r|r|} \hline \hline
Source &Uncertainty in MeV& Correlation coefficient  \\
 &&  with previous results \\\hline
\hline
\bf Experimental uncertainties \rm&&\\
\hline
W Statistics &39& 0\\  
\hline
Electron response model& 33&0\\  
Electron resolution model& 10&0\\
Hadronic recoil model& 41&0\\
Electron efficiencies & 19&0\\
Backgrounds   & 6&0\\
\hline
\bf Production uncertainties\rm &&\\
\hline
PDFs& 20& 1.0\\
EWK radiative corrections & 7 &1.0\\
Boson $p_T$  & 1&0\\
$M_W$ & 5 &1.0\\ \hline
\end{tabular}
\caption{Contributions (in MeV) to the uncertainty on the measurement of \GW\ in  D0 data from Run-II . \label{table:err}}
\end{center}
\end{table}

\section{Correlation of the D0 Run II result with other measurements}

Experimental  uncertainties on the new D0 measurement of \GW\  are dominated by 
the statistical uncertainty on the number of  $W$ events found in the high mass region sensitive to the width,  and by uncertainties in the  energy response of the D0  detector. Energy response functions are derived from events containing $Z$ bosons and their uncertainties are almost purely statistical. All of the experimental uncertainties in the new D0 measurement of \GW\ are assumed to be uncorrelated
with previous measurements.

Three systematic uncertainties from the production of  $W$ and $Z$ bosons are assumed to be fully correlated among all
Tevatron measurements, namely (i)
 the parton distribution functions (PDFs),
(ii)  the mass of the $W$ boson ($\MW$) and
(iii) the electroweak radiative corrections (EWK RC).

The D0 measurement also   includes an uncertainty in the  models  of the $W$ and $Z$ boson $p_T$ distributions,  which is derived from a global fit to deep-inelastic scattering and hadron collider data~\cite{ref:Landry}.  In previous analyses, this source of uncertainty is treated differently, and it is therefore regarded as uncorrelated with
the earlier measurements.
 
Current estimates of  uncertainties from  radiative corrections include a significant statistical component. The WGRAD/ZGRAD~\cite{ref:WGRAD} and PHOTOS~\cite{ref:PHOTOS} models are used in the different measurements and yield results consistent within the  statistical uncertainties. We assume that the  effects of radiative corrections are 100\% correlated between  all measurements because the models used are very similar.

\section{Combination of Widths of the W Boson}\label{combo}

\begin{table}[htbp]
\begin{center}
\renewcommand{\arraystretch}{1.30}
\begin{tabular}{| l | r || r | r || r | r | }
\hline       
       &  \multicolumn{3}{|c||}{{\RunI}} 
       &  \multicolumn{2}{|c|}{{\RunII}} \\
\hline 
 
       & \multicolumn{1}{|c|}{ CDF-Ia }
       & \multicolumn{1}{|c|}{ CDF-Ib }
       & \multicolumn{1}{|c||}{ D0-Ib}
       & \multicolumn{1}{|c|}{ CDF } 
       & \multicolumn{1}{|c|}{ D0 }  \\
\hline       
\hline 			       
\GW\ (published)                                           & 2,110      & 2,042.5      &  2,231    & 2,032     & 2,028.3 \\
Total uncertainty (published)                                   &  329      &  138.3       &  172.8     & 72.4     & 72 \\
\MW\ used in publication                             & 80,140    & 80,400    & 80,436  &  80,403   & 80,419 \\ 
Correction to \GW\ from \MW\                             & $-$78    & 0.3         &  11.1    &   1.2       &   6.0 \\
\GW\ (corrected)                                          & 2,032     & 2,042.8  & 2,242.1  & 2,033.2    &  2,034.3 \\ \hline   
Total  uncertainty(corrected)                        & 329.3    & 138.3    & 172.4   &    72.4       & 71.9 \\ 
Uncorrelated  uncertainty (corrected) & 327.6        &	136.8   & 167.4  &  68.7	    & 68.5 \\ \hline \hline
PDF uncertainty(published)                       &    0      &    15      &    39     &     20       &    20  \\
PDF uncertainty (this analysis)                   &    15      &    15      &    39     &     20      &    20  \\
EWK RC  uncertainty   &    28      &    10     &     10    &    6          &    7 \\ 
\MW\ uncertainty (published)                      &    0      &    10      &    15     &     9     &   5  \\
\MW\ uncertainty  (this analysis)  &7&	7&	7&	7	&7\\
\MW\ extrapolation                 
&26	&0	&4&	0	&2\\  \hline 
\end{tabular}
\end{center}
\caption[Input measurements]{Summary of the five measurements of
$\GW$ performed by CDF and D0. All numbers are in $\MeV$.  The
published values and the corrected values (assuming $M_W$ is the 
2009 world average of
80,399 $\pm$   23  MeV) used in the average are
shown. The three sources of correlated systematic uncertainty (PDF, EWK RC,
\MW) are given  explicitly.}
\label{tab:GW-inputs}
\end{table}

\subsection{Corrections for changes in  \MW}
As in the case of the combined  mass analysis, we have applied  corrections to achieve
consistency across all  input results. The CDF Run-Ib results have been
recombined using the BLUE method,  and all results  are corrected
so that  \GW\ is evaluated assuming the world-averaged (December 2009) mean value of $\MW=$ 80,399 $\pm\ 23$ MeV~\cite{LEPEWWG}.
We correct for  the $\MW$ assumptions in the initial publications using the relation $\Delta\GW = (-0.3 \pm 0.1) \times
\Delta\MW$. This is the average of the shift in \GW\  empirically determined by CDF and D0 when \MW\ is varied.  We include an  uncertainty  of 0.1 $\Delta\MW$, for this correction. In addition, we have re-evaluated the uncertainties on \GW\  due to the uncertainty in \MW. The world average uncertainty of 23 MeV in \MW\ yields an uncertainty in \GW\ of 7 MeV which replaces the \MW\ uncertainty assumed  in the original publications.
  In most cases, our improved knowledge of \MW\ has decreased the estimated uncertainty from the input \MW.


\section{Results}

The combined Tevatron value for \GW\ is:
\begin{eqnarray*}
\GW & = & 2,046 \pm 49~\MeV\\
\end{eqnarray*}

\begin{table}[!hbp]
\begin{center}
\begin{tabular}{|c |r|}  \hline 
           &    Relative Weights in \%  \\ \hline \hline
CDF Ia     &   1.6  \\
CDF Ib     &   10.8  \\
D0 I     &  5.4 \\
CDF II     &  41.2  \\ 
D0 II    &  41.2 \\ \hline \hline
\end{tabular}
\caption{Relative weights of the individual contributions in \%. \label{contribution}}
\end{center}
\end{table}

\begin{figure}[bp]
\begin{center}
\includegraphics[width=0.8\textwidth]{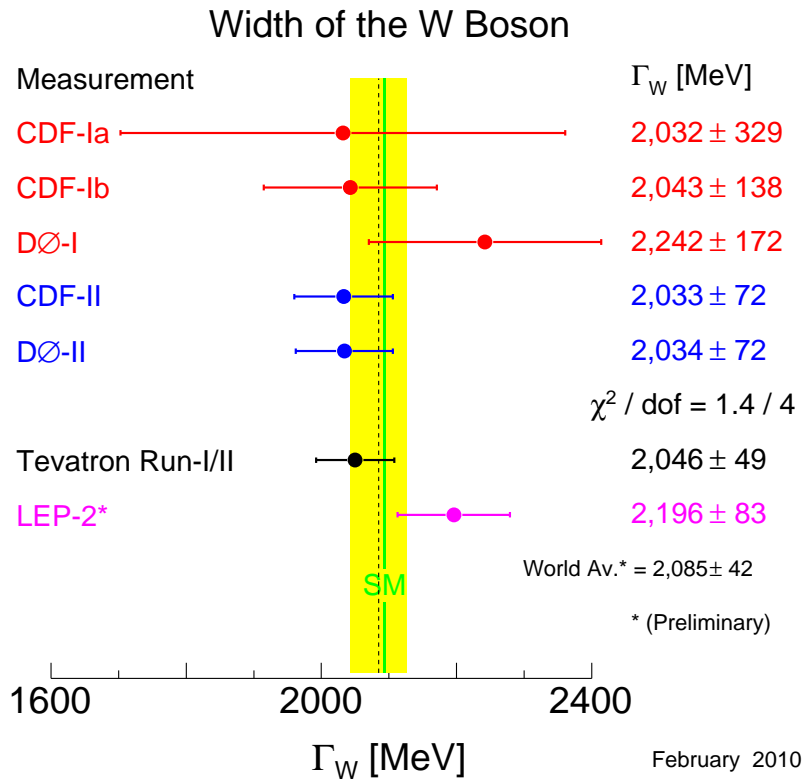}
\end{center}
\caption[Comparison of   measurements of the width of the $W$-boson.]
{Comparison of   measurements of the width of the  $W$-boson and their
average.  The most recent preliminary result from
LEP-2~\cite{LEPEWWG} and the Standard Model prediction are also
shown. The  Tevatron values  are corrected for small inconsistencies in theoretical assumptions among the original publications .}
\label{fig:gw-bar-chart} 
\end{figure}

The combined Tevatron result has a $\chi^2$ of 1.4 for 4
degrees of freedom, corresponding to a probability of 84\%. All measurements are in good agreement with each other as can be seen 
  in Figure~\ref{fig:gw-bar-chart}, where the individual results and this combination are shown.

The total uncertainty on the combined Tevatron \GW\ is 49 MeV  and consists of  the following components: an uncorrelated uncertainty  of 
44 MeV and correlated systematic   contributions  from parton distribution functions of  20 MeV, 
electroweak radiative corrections of 7.4 MeV, and input  $W$-boson mass of 7.4 MeV, for a total correlated systematic 
uncertainty  of 23 MeV. 
The global correlation matrix of  the five Tevatron  measurements is shown in Table~\ref{tab:GW-corr}.

\begin{table}[htbp]
\begin{center}
\renewcommand{\arraystretch}{1.30}
\begin{tabular}{|l||c|c|c||c|c|}
\hline       
       &  \multicolumn{3}{|c||}{{\RunI}} 
       &  \multicolumn{2}{|c|}{{\RunII}} \\
\hline
       & \multicolumn{1}{|c|}{ CDF-Ia }
       & \multicolumn{1}{|c|}{ CDF-Ib }
       & \multicolumn{1}{|c||}{ D0-Ib}
       & \multicolumn{1}{|c|}{ CDF } 
       & \multicolumn{1}{|c|}{ D0 }  \\
\hline       
\hline 			       
CDF-Ia      &  1.00      &  0.02     &     0.02      &   0.03     &  0.03 \\
CDF-Ib       &            &   1.00      &     0.03      &   0.04      &  0.04  \\
D0-I          &            &              &     1.00        &   0.07      &  0.07  \\
CDF-II        &            &              &                  &   1.00        &  0.09  \\
D0-II        &             &              &                   &                &  1.00  \\ \hline
\end{tabular}
\end{center}
\caption[Input measurements]{Matrix of global correlation coefficients among the 5 measurements
of Table~\ref{tab:GW-inputs}.}
\label{tab:GW-corr}
\end{table}

 A combination with the latest LEP-2 average value,  $\GW =$ 2,196 $\pm$  83~MeV~\cite{LEPEWWG}, assuming no correlation between the Tevatron and LEP-2 measurements, gives a preliminary 
world average of $\GW =$ 2,085 $\pm$ 42 ~MeV with a $\chi^2$ of 2.4 for  one degree of freedom.  This world average value is in  agreement with the SM 
prediction of $\GW = $ 2,093 $ \pm$ 2~MeV~\cite{PR}.

\section{Summary}

Combinations of the direct CDF and D0 measurements of the 
total decay width of the W boson are presented. Corrections have
been made to achieve a consistent treatment across published Tevatron
measurements, corrected for inconsistencies in Standard Model parameters. The Tevatron average  result is 
$\GW=$ 2,046 $\pm$ 49~ MeV, and a preliminary world average including both the Tevatron and LEP2 is $\GW= $ 2,085 $\pm$ 42~\MeV.

\end{document}